\documentclass[12pt]{emulateapj}
\newcommand{\beq}{\begin{equation}}
\newcommand{\eeq}{\end{equation}}

\newcommand{\hi}{H{\sc i}~}

\newcommand{\hia}{H{\sc i}}

\newcommand{\citei}[1]{\citeauthor{#1} \citeyear{#1}}	

\newcommand{\kms}{km ${\rm s^{-1}}$~}
\newcommand{\kmsa}{km ${\rm s^{-1}}$}
\newcommand{\hipstar}{HIP\,47513}
\newcommand{\nai}{Na\,{\sc i}}

\usepackage{color}

\begin{document}

\title{The Local Leo Cold Cloud and New Limits on a Local Hot Bubble}

\author{J.~E.~G.~Peek\altaffilmark{1}}
\author{Carl Heiles\altaffilmark{2}}
\author{Kathryn M. G. Peek\altaffilmark{2}}
\author{David M. Meyer\altaffilmark{3}}
\author{J. T. Lauroesch\altaffilmark{4}}
\altaffiltext{1}{Department of Astronomy, Columbia University, Pupin Physics Laboratories, 550 West 120th Street, New York, New York, 10027. goldston@gmail.com}
\altaffiltext{2}{Department of Astronomy, University of California, Berkeley, Berkeley, CA 94720}
\altaffiltext{3}{Center for Interdisciplinary Exploration and Research in Astrophysics, Department of Physics and Astronomy, Northwestern University, Evanston, IL  60208}
\altaffiltext{4}{Department of Physics and Astronomy, University of Louisville, Louisville, KY  40292}
\slugcomment{Accepted for Publication in the Astrophysical Journal}

\begin{abstract}
We present a multi-wavelength study of the local Leo cold cloud (LLCC), a very nearby, very cold cloud in the interstellar medium. Through stellar absorption studies we find that the LLCC is between 11.3 pc and 24.3 pc away, making it the closest known cold neutral medium cloud and well within the boundaries of the local cavity. Observations of the cloud in the 21-cm \hi line reveal that the LLCC is very cold, with temperatures ranging from 15 K to 30 K, and is best fit with a model composed of two colliding components. The cloud has associated 100 micron thermal dust emission, pointing to a somewhat low dust-to-gas ratio of 48 $\times$10$^{-22}$ MJy sr$^{-1}$ cm$^{2}$. We find that the LLCC is too far away to be generated by the collision among the nearby complex of local interstellar clouds, but that the small relative velocities indicate that the LLCC is somehow related to these clouds. We use the LLCC to conduct a shadowing experiment in 1/4 keV X-rays, allowing us to differentiate between different possible origins for the observed soft X-ray background. We find that a local hot bubble model alone cannot account for the low-latitude soft X-ray background, but that isotropic emission from solar wind charge exchange does reproduce our data. In a combined local hot bubble and solar wind charge exchange scenario, we rule out emission from a local hot bubble with an 1/4 keV emissivity greater than 1.1 Snowdens / pc at 3 $\sigma$, 4 times lower than previous estimates. This result dramatically changes our perspective on our local interstellar medium.

\end{abstract}
\section{Introduction}\label{intro}
The Galactic interstellar medium (ISM) is the pervasive gas that fills our Galaxy. Our ISM has a huge range of temperatures and densities, from molecular gas at densities above $10^4$ cm$^{-3}$ and temperatures of only 10 K, to the hot ionized medium (HIM), which can exceed 10$^6$ K in temperature, with densities dropping below 10$^{-2}$ cm$^{-2}$. The dynamic interplay between these different phases of the interstellar medium is not very well understood: How is the cold neutral medium (CNM) formed? What kind of gas fills low-density volumes in the ISM? How is it created? One way to approach these questions is to examine the ISM nearest the sun, an area called the local cavity, where our observations can most easily disentangle individual clouds, and examine their formation, structure, interaction, and destruction. Indeed, the canonical basis for our understanding of the interplay between hot and cold ISM phases, \citet{MO77}, takes observations of the local cavity as its starting point.

Up until recently, the only ISM clouds thought to exist near the sun were the complex of local interstellar clouds (CLIC), a group of about 15 low column density ($N_H \sim 10^{18}$ cm$^{-2}$), partially ionized clouds within about 15 pc of the sun (e.\ g.\ \citei{ RL08}; RL08). Recent observations by \citet{Meyer:2006dn} have added an entirely new animal to the local menagerie: a very cold CNM cloud, which we call the Local Leo Cold Cloud (LLCC). 

The first observation of the LLCC was conducted by \citet{Verschuur69}, who discovered a very cold component of the ISM while looking towards intermediate velocity clouds. The very narrow observed linewidth in the 21-cm hyperfine transition of hydrogen led to his determination that this parcel of ISM must have a temperature of less than 30 K. \citet{KV72} followed this work up by mapping the LLCC (roughly $\alpha =$ 10 h, $\delta =$ 10$^\circ$), showing it to be composed of two angularly distinct components, each a few square degrees in area. \citet{CK80} found a spin temperature of about 20 K using the \hi absorption-line towards the quasars 3C 225a and 3C 225b, but found no evidence for molecules in the 1665 and 1667 MHz line of hydroxyl (OH). The LLCC was later independently discovered by \citet{HT03}, who saw it again in absorption toward 3 sources in their Arecibo millennium survey, and found temperatures of 14, 17 and 22 K. \citet{Meyer:2006dn} observed stars toward this set of clouds and found that every star in the direction of the cloud showed significant Na {\sc I} absorption at exactly the cloud's velocity in \hia, putting an upper limit of the distance to the cloud of 42 pc (HD 83683). Most recently \citeauthor{haud10} (\citeyear{haud10}; Haud10)  found the LLCC  in the Leiden-Argentina-Bonn survey data, and found that it was connected to a much larger ribbon of clouds with consistent linewidths and a continuous velocity distribution, stretching from -10$^\circ$ to +40$^\circ$ declination and from 7h to 11h right ascension. This local ribbon of cold clouds (LRCC) extends through the constellations Sextans, Leo, Cancer, and Lynx, with small pieces in Hydra, Gemini, and Auriga.

The discovery that there is an extended group of nearby CNM clouds is intriguing in the context of the local cavity. The sun lives roughly in the middle of a largely evacuated volume of gas, roughly 200 pc wide in the Galactic plane, and perhaps somewhat prolate (or even open) towards high Galactic latitude \citep[e.g.,][]{Vergely10, Welsh10}. The edges of this volume of gas are typically defined by the distance at which the total hydrogen column density reaches $10^{19.3}$ cm$^{-2}$. The cavity seems to have a rather discrete edge, with very little material inside it. Since the original Wisconsin rocket sounding experiments \citep[e.g.,][]{Burstein77} it had been proposed that this cavity is filled with an approximately million-degree plasma, emitting thermally in soft X-rays. 
The million-degree plasma would explain the observed soft X-rays, which would be absorbed by neutral gas outside the cavity if its source were further away. Morphologically this made good sense as well; the bubble could have been carved out by a series of supernova explosions and stellar winds, which would have supplied the energy to heat the gas \citep[e.g.,][]{CR87}. This is in excellent agreement with theories of the ISM as a whole, with supernovae peppering the disk, blowing bubbles and launching material above the Galactic plane \citep[e.g.,][]{deAvillez00}. Once the ROSAT all-sky survey soft X-ray background \citep[RASS; SXRB;][]{Snowden97} maps were available, \citet{KS2000} proposed a modified mechanism to explain the soft X-ray signature; part of the diffuse soft X-ray emission came from the halo and was partially absorbed by low column, high-latitude \hia, and part of it originated from the local hot bubble. 

This picture has been challenged by a number of different observations and theories, including concerns regarding O {\sc vi} interstellar absorption and the relationship between cavity depth and soft X-ray flux (see \citei{WS10} for a detailed summary), but none so concerning as the solar wind charge exchange (SWCX), a process that may be an alternate origin for the SXRB. ROSAT observations of comet Hyakutake showed significant, unexplained X-rays, which were later determined to be from the electronic cascade of exchanged electrons from the relatively neutral comet to highly ionized solar wind metals \citep{Lisse96}. Both \citet{Cox98} and \citet{Freyberg98} suggested that this same SWCX process could generate X-rays in the heliopause, where the ionized solar wind met the largely neutral ISM of the local interstellar cloud (LIC). Accurate models of the SWCX emission are very difficult to generate, as the efficiency of exchange is largely unknown and the atomic physics of high ions is not fully constrained; as such there is no clear consensus as to how much of the soft X-ray emission could be coming from SWCX, rather than a LHB. That being said, models by \citet{RC03} and \citet{Koutroumpa09} show that somewhere between 50\% and 100\% of the X-rays could be originating from SWCX, which leaves the fate of the LHB X-rays very uncertain.

It has been suggested by various authors \citep[e.g.,][]{CR87, WS10} that an excellent way to determine the provenance of the soft X-rays would be to find a nearby cloud capable of providing a screen to soft X-rays. If a shadow for this cloud could be found, we could distinguish between X-rays coming from a LHB and X-rays coming from SWCX. We believe that the LLCC is just such a cloud, at the right distance and column density, and we devote a large fraction of this paper to carefully determining the crucial parameters needed to use the LLCC in this proposed shadowing experiment.

In Section \ref{obsdatred} we describe the observation and data reduction of a multi-wavelength data set, including new observations in the optical and radio, as well as archival data in the infrared and X-ray. In Section \ref{analysis} we discuss how we analyze these data in the context of the LLCC. In Section \ref{cloudprop} we discuss various properties of the cloud we can determine from the observations. In Section \ref{ilc} we discuss the implications of the LLCC observations for the contents of the local cavity, including the relationship to CLIC clouds and the implications for the LHB and SWCX theories. We conclude in Section \ref{conc}.

\section{Observations and Data Reduction}\label{obsdatred}
One of the main advantages of studying such a nearby structure is its scale on the sky; Haud10 detects the LRCC over 86 square degrees, and we find the LLCC alone to cover 22 square degrees. Because of this large area, we are capable of detailed morphological studies with low resolution observations. In particular, we can use archival all-sky surveys to pin down the structure of the object over a broad range of wavelengths. Additionally, the large area allows us to study the object more easily in absorption as more background sources are available.

\subsection{Radio: GALFA-HI}
The LLCC was originally discovered in the 21-cm (1420 MHz) hyperfine transition of \hia, and this kind of observation remains the best way to study the morphology of the structure. Temperature measurements of $\sim$20 K from \citet{CK80} and \citet{HT03} indicate that the cloud should be largely neutral. Measured column densities below $10^{21}$ cm$^{-2}$ and the non-detection of OH \citep{CK80} indicate that the cloud should be largely atomic and not entirely optically thick to \hia. Thus, \hi is an excellent tracer of the bulk of  the cloud.

We observed the LLCC as part of the Galactic Arecibo L-Band Feed Array \hi (GALFA-\hia) survey, an ongoing survey of \hi conducted with the Arecibo 305m telescope using the Arecibo L-band Feed Array (ALFA). The GALFA-\hi survey is a survey of all Galactic \hi within the Arecibo observing range ($-1^\circ < \delta < 38^\circ$), at a spectral resolution of 0.184 \kmsa, and an angular resolution near 4$^\prime$. The details of the observation methods, data reduction, and properties of the GALFA-\hi survey can be found in \citet{Peek11}. Observations for this region were conducted both as part of a targeted map, and as part of the Turn On GALFA Survey, a commensal (simultaneous) survey conducted alongside the the Arecibo legacy fast ALFA project. The targeted map was conducted in a high-speed, basketweave or meridian-nodding mode, whereas the Turn On GALFA Survey data were collected in a low-speed drift mode. Most areas of the map were covered by both modes, and have an RMS noise of 0.17 K over a 0.184 \kms channel, but a few areas were only covered with the basketweave mode, and have an RMS noise of 0.36 K per 0.184 \kms channel. It is important to point out that the detailed analysis of the \hi column density found in \S \ref{ghi_analysis} could not be accomplished without the exquisite spectral resolution of the GALFA-\hi survey. We note that while we refer to the section of cloud we are examining as the Local Leo Cold Cloud, the southern component of the LLCC is actually in Sextans. 

\subsection{Infrared: IRAS \& IRIS}
A key result from the \emph{Infrared Astronomical Satellite (IRAS)} was the discovery of the correlation between the infrared diffuse emission and the Galactic \hi column density for $|b| > 10^\circ$ (\citei{Low84}). The conclusion reached by these authors, and others, is that there is a relatively uniform interstellar radiation field in the Galaxy and a relatively constant large-grain dust fraction in the diffuse \hia. The interstellar radiation field heats the dust, which reradiates in the infrared, which in turn creates the observed correlation between \hi column density and dust emission. Thus, the investigation of the large-grain dust contents of the LLCC can be conducted by examining the infrared flux in the IRAS data. IRAS observed the vast bulk of the sky for 300 days in 1983, in the infrared wavelength bands centered on 12, 25, 60, and 100 $\mu$m \citep{Beichman87}. These data have been reduced in numerous ways and have produced many different maps: the original SkyFlux Atlas, the IRAS Sky Survey Atlas, the maps produced in \citet{SFD98}, and the IRIS maps described in \citet{M-DL06}. We use these most recent IRIS maps, as they are produced at 4$^\prime$ resolution, nearly identical to the resolution of the GALFA-\hi maps. In addition to the IRIS reduction, we further clean the data set by removing point sources down to a magnitude 1 Jy. We remove these point sources by fitting with Gaussian PSF, and excise any point sources we cannot cleanly remove. We note that the presence of IRC+10216 in this field has made the analysis of the 60, 25, and 12 micron band impossible, as this exceedingly bright star has contaminated a large swath of the map. 

\subsection{Optical: KPNO \& CPS}\label{optical}

Based on observations of the interstellar \nai\ D$_2$ $\lambda$5889.951
and D$_1$ $\lambda$5895.924 absorption toward 33 stars in the LLCC sky
region, \citet{Meyer:2006dn} were able to place a significant upper
limit on the LLCC distance.  Using the updated {\it Hipparcos}
parallaxes of \citet{vanLeeuwen07}, this upper limit corresponds to the
distances (39.9$^{+0.4}_{-0.3}$ and 40.5$^{+1.1}_{-1.0}$ pc) of the
nearest stars (HD 83808 and HD 83683) exhibiting LLCC \nai\ absorption.
The challenge in tightening this distance constraint further is
finding nearer stars within the LLCC \hi 21-cm emission sky envelope
that are bright enough for high-resolution optical absorption-line
spectroscopy.  Given the paucity of such stars, another possibility is
to consider nearer stars just outside the \hi envelope that might be
bright enough to detect very weak optical absorption at a gas column
threshold more sensitive than the 21-cm observations.

Fortunately, there is a very bright (V~$=$~1.35) star, Regulus ($\alpha$~Leo),
at a distance of 24.3~$\pm$~0.2 pc \citep{vanLeeuwen07} and a sky
position that is 3$\arcdeg$ outside the LLCC 21-cm envelope.
\citet{Meyer:2006dn} obtained high S/N spectra of Regulus in
the \nai\ wavelength region to divide out the atmospheric absorption
lines in the spectra of their program stars.  These Regulus
spectra reveal no evidence of any interstellar \nai\ absorption to
high precision.  Since Regulus is often used as a spectral
standard for observations in other wavelength regions, we examined
our archive of data obtained with the 0.9 m coud\'e feed telescope and
spectrograph at Kitt Peak National Observatory (KPNO) over the past
fifteen years.  Observations of Regulus in the vicinity of the
interstellar Ca II K $\lambda$3933.663 and H $\lambda$3968.468 lines
were obtained with this instrumentation in November 2002 in support
of a study of variable interstellar absorption toward $\rho$~Leo
\citep{LM03}.  Eight F3KB CCD echelle spectra spanning a total exposure time of
80 minutes were taken of Regulus with a
spectrograph configuration yielding a measured velocity resolution
of 3.6~km~s$^{-1}$ at the Ca II K wavelength.  We utilized the NOAO
IRAF echelle data reduction package to bias-correct, scattered light--correct, flat-field, wavelength-calibrate, order-extract, and sum the
individual CCD exposures into the final net Regulus spectrum.

For the lower limit on the LLCC distance, we borrowed archival data from
the California Planet Search of the nearby M-dwarf \hipstar.
This star is at a distance of $11.26 \pm 0.21$ pc \citep{vanLeeuwen07}.
In addition to \hipstar, we used spectra of four additional 
M-dwarfs with similar features as comparison stars.
Those other spectra allow us to compare the \nai \ D$_{1}$ line core
of \hipstar\ with the cores of other M-dwarfs, and thus to place a 
stronger constraint on the interstellar \nai\ column along the \hipstar\ line of 
sight.

Our five M-dwarf spectra were collected with HIRES, the 
high-resolution echelle spectrograph on Keck I \citep{vogt94}, 
between 2005 February and 2006 December. 
Details are in Table \ref{keck_table}. 
The spectra were originally collected with the intention of detecting
extrasolar planets; see \citealt{marcy92} for a full description 
of the CPS and its goals.
Because of their radial velocity measurement technique, the vast 
CPS archive contained only a few useful spectra of each star.
The CPS places an iodine cell in the light path of its stars, which 
creates a forest of I$_{2}$ absorption lines for radial velocity reference.
Here we make use only of the iodine-free ``template'' spectra, which
contain no I$_{2}$.

The spectra have a resolution $R = 70,000$ at $\lambda=5500$\AA. 
We chose the stars in Table \ref{keck_table} because they were M-dwarfs
with $B-V$ within $\pm 0.12$ of \hipstar, and they had iodine-free 
observations in the CPS archive. 
We began with 33 candidate stars that fit those two criteria, 
then inspected by eye three separate segments of the spectra, 
away from the \nai D lines.
We graded candidates in how well their spectral features matched 
the features of \hipstar. 
Four spectra were considered excellent matches, and those are the four
we chose to include in this analysis.

\begin{deluxetable}{lrrrr}
\tablecolumns{4}
\tablecaption{M-dwarfs from the California Planet Search included in this study.\label{keck_table}}
\normalsize
\tablewidth{0pt}
\tablehead{  \colhead{Star}  & \colhead{$V$}  & \colhead{$B-V$} & \colhead{Date} & \colhead{Exp.\,($s$)}}
 \startdata
GL 250 B        & 10.05      & 1.42      & 2006 Jan & 500 \\ 
\bf{HIP 47513}   & 10.38      & 1.49      & 2006 Dec & 600\\ 
HD 97101 B      & 9.95       & 1.57      & 2005 Feb & 500 \\ 
HIP 70865       & 10.68      & 1.47      & 2006 Jan & 500 \\ 
HIP 115562      & 10.05      & 1.47      & 2006 Jan & 600  
\enddata
\end{deluxetable}

\subsection{X-ray: ROSAT}
We examine X-ray data in order to determine the provenance of the SXRB via the X-ray shadow of the LLCC (see \S \ref{intro}). The \emph{R\"ontgensatellit} (ROSAT; \citei{Truemper92}) produced a flood of high-quality, all-sky data taken during 6 months in 1990 and 1991. Among these data were the soft X-ray background (SXRB) maps. We use the second data release of these maps, refined from the original reduction to have higher resolution (12$^\prime$) and fidelity \citep{Snowden97}. The SXRB maps cover 98\% of the sky in 3 wavebands: 1/4 keV, 3/4 keV, and 1.5 keV. We focus only on the 1/4 keV (C-band) SXRB maps in this work, as \hi can act as very effective absorber of photons at this energy level, even at the intermediate column densities of the LLCC. The SXRB maps were reduced so as to exclude variable local emission and the contribution of point sources. The brightness of the SXRB maps are reported in units of $10^{-6}$ counts s$^{-1}$ arcmin$^{-2}$ (Snowdens). 

\section{Analysis}\label{analysis}

\subsection{Radio: GALFA-HI}\label{ghi_analysis}

\subsubsection{HI Profile Saturation and Optical Depth}

\citet{HT03} measured \hi absorption spectra and derived \hi excitation
temperatures $T_x$ for the strong radio continuum sources 3C225a,
3C225b, and 3C237, and found $T_{x} = 21.53 \pm1.30, 17.44 \pm 1.80, 13.61 \pm 0.26$ K, respectively. These three low temperatures suggest that the large parts of the LLCC are very cold,
$\lesssim 20$ K, throughout.  Given that we find brightness temperatures in this range for the center of the cloud,  we expect large parts of the cloud to be optically thick at line center, i.e., saturated. The
derivation of accurate \hi column densities requires including this line saturation effect; if we neglect the saturation, then the derived values are too small.

It is easy to correct for line saturation at the position of a
sufficiently strong continuum source, because then we can measure not
only the emission profile but also the optical depth profile. However,
it is difficult to correct for saturation throughout the cloud's area,
because without a background continuum source our only indication of
line saturation is the line shape. 

Knapp \& Verschuur (1972) attacked this problem by assuming that the
intrinsic line shape is accurately represented by a single saturated
Gaussian component and used least-squares fits to determine its degree
of saturation and spin temperature $T_x$. However, we have found that the stronger \hi
profiles tend to be double-peaked, so this model is invalid.

\begin{figure}
\begin{center}
\includegraphics[scale=0.4]{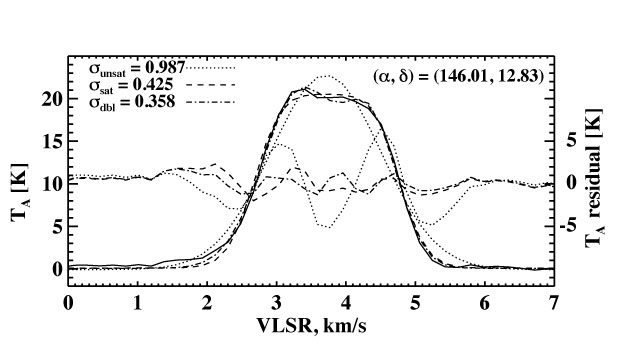}

\caption{A comparison of a single-component optically thin (dotted), optically thick (dashed), and a double-component optically thin (dash-dotted) fit of the \hi line profile towards a typical, high-column part of the LLCC (data are shown by the solid line). Residuals between the fit and data are overplotted with scale at right. It is immediately clear that an unsaturated Gaussian fit does a very poor job at fitting the line shape, and that even a saturated Gaussian has trouble reproducing the double-peaked profile in the data. While the double-component optically thin fit largely succeeds, it ignores the effects of \hi optical depth, and thus is unphysical. 
  \label{satplotfig}}
\end{center}
\end{figure}

Figure \ref{satplotfig} shows an example spectrum towards the core of the cloud. Visually, the lines look double. This is confirmed by least-squares fitting of three models. The
solid line is the observed profile, the dotted line the single
unsaturated Gaussian fit, the dashed line the saturated Gaussian fit,
and the dash-dot line the double unsaturated Gaussian fit. The
dispersions of the data from each fit are indicated on the upper left of
each plot. The double unsaturated Gaussian fit is the best fit, and we find this to be generally true across the highest column parts of the cloud. We must therefore consider multi-component models to accurately fit these data.

\subsubsection{Five Fitting Models}

The best fit in Figure \ref{satplotfig}, the double unsaturated
Gaussian, neglects the effects of \hi optical depth, and will therefore not yield physically reasonable quantities. To explore the optical depth of the cloud and its behavior over the entire cloud area, we fit five models to
each pixel's \hi profile: \begin{description}

\item[(1)] A single unsaturated Gaussian. This is appropriate for pixels with
  weak, single-component lines. This model has three unknown parameters: the height, center, and width of the Gaussian.

\item[(2)] A double unsaturated Gaussian. This is appropriate for pixels with
  weak, double-component lines. This model has six unknown parameters:
  the heights, centers, and widths of the two Gaussians. 

\item[(3)] A double Gaussian in which the spin temperatures are equal. We consider this our most
  ``physically reasonable'' model; \hi temperatures are
  determined by microphysical heating and cooling processes that are not
  expected to change rapidly with position, and thus we do not expect the two components to have different spin temperatures. This model has seven
  parameters: the spin temperature and the central optical depths,
  centers, and widths of the two Gaussians.

\item[(4)] A double Gaussian in which each component has its own spin
  temperature, central optical depth, center, and width, for a total of eight
  parameters between them. An additional (ninth) parameter is the line-of-sight
  ordering of the two components, i.e.,\ which lies closer to the
  observer (and thus absorbs the radiation from its more distant
  sibling).

\item[(5)] A double Gaussian in which the more distant is optically thin, and
  the nearer component is optically thick (and thus absorbs the
  radiation from its more distant sibling). This model has seven
  parameters. 
\end{description}

\subsubsection{Procedure for Fitting Gaussians}\label{fitgau}

Fitting Gaussians is a difficult and non-unique process because the
functions are nonlinear with respect to the fitted parameters. Such fits
require guesses; the equations are linearized in a multidimensional
Taylor series around the guesses and then solved to find corrections to the
guesses. The corrections are applied and the process is iteratively
repeated until convergence or failure. The final results depend on the
guesses, which makes the whole process subjective. See the Appendix for a discussion of the guessing procedure. \footnote{There are
  schemes to explore the multidimensional parameter space for the
  guesses, but they are computationally intensive and not perfectly
  reliable.}

The fit for model (1) always converges. The fits for the other models
don't always converge. If a fit does converge, we record the mean error
of the fitted points, which we denote $\sigma$, and also the mean error
of points within the velocity ranges where the fitted intensity exceeds
0.1 of the peak fitted intensity, which we denote $\sigma_{line}$. 
Fits for model (3) usually don't converge because for the more
distant component the only handle on its optical depth is its line
shape. We find this is too weak a handle, particularly in the presence of
noise and the line shape modification produced by the nearer
component. For this reason, we don't include model (3) in our subsequent
discussion.

\subsubsection{Model 4 or Model 5?}

Our goal is to settle on three models that apply to the whole cloud. Two
of these are the single- and double-component optically thin models,
which are appropriate for weak lines, where the effects of optical depth have an undetectable effect on our observed line shape. These are models (1) and (2). The third, for strong lines, is either model (4) or
model (5). We choose between these by examining their $\sigma_{line}$
statistics. 

Here we compare $\sigma_{line}$ values for various
model pairs. For most of the
fits model (2) has a lower $\sigma_{line}$ than model (1) indicating that the vast majority of sightlines have double components. 

By doing the same comparison between model (4) versus model (5), we find the vast majority of sightlines are better fit by model (5). Philosophically, we would
have much preferred the physically motivated model (4), but we must be
driven by the data. A second philosophical objection is illusory: why
should the near side be optically thick and the far be optically thin?
The answer is that the far side may well be optically thick; we cannot
derive its optical depth because, as discussed in the context of model (3), all we have is its
line shape, and this is an insufficient ``handle.''

We also make the same comparison between model (5) and model (2), i.e., including optical depth effects versus
assuming optically thin components. We find that while a majority of points have lower $\sigma_{line}$ for model (5), there remain a large minority of sight lines that clearly prefer model (2). This is to say that optically thin fits are useful along with the optically thick fits. 

\begin{figure}
\begin{center}
\includegraphics[scale=0.3]{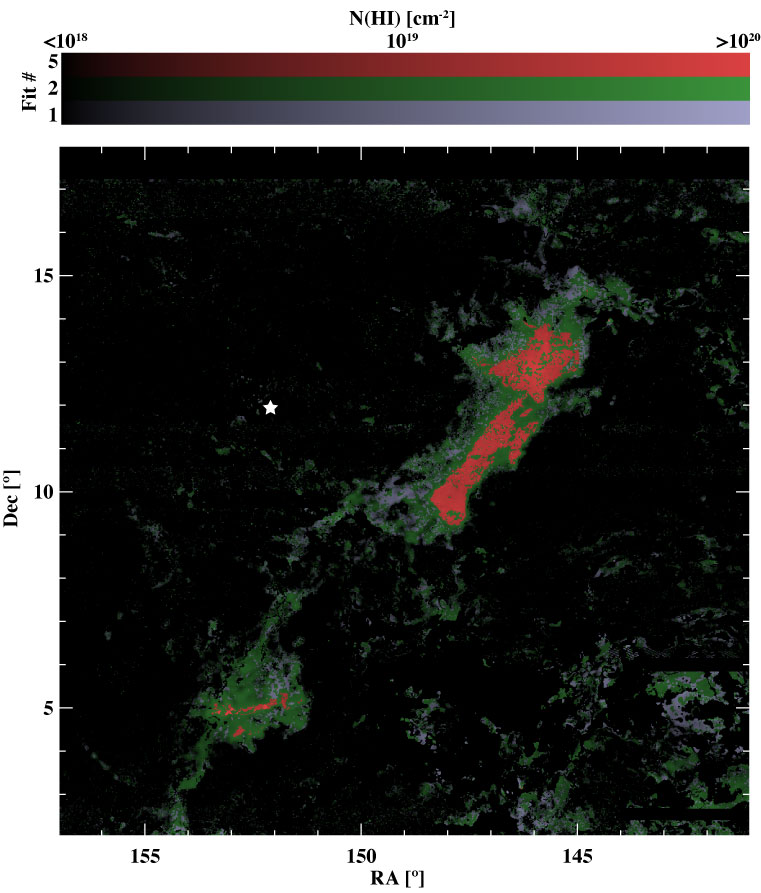}
\end{center}
\caption{Brightness is $\propto log_{10}N(HI)$, and color indicates the model chosen for each pixel: model 1 is represented in blue, model 2 in green, and model 5 in red. The highest column regions are typically best fit with model 5, and the lowest column regions are best fit with model 1, with intermediate column densities best fit by model 2. The position of Regulus is marked by a white star.
 \label{img_fitnr}}
\end{figure}

For each pixel, we choose the converged model with the lowest
$\sigma_{line}$. Figure \ref{img_fitnr} shows an image of the cold cloud
in which the brightness indicates column density and the color shows the
model selected. For each pixel, the column density is calculated from
the selected model.

\subsection{Infrared}
\begin{figure}
\begin{center}
\includegraphics[scale=0.4, angle=0]{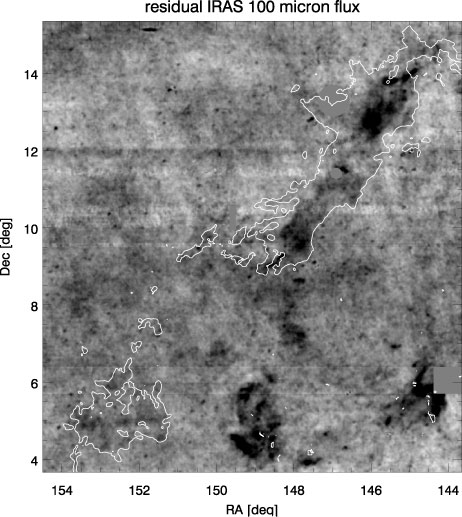}
\caption{The residual 100 $\mu$m flux from the IRIS maps toward the LLCC, shown on a range from -0.7 to 0.7 MJy sr$^{-1}$. The LLCC $10^{19}$ cm$^{-2}$ \hi contour is overlaid in white to guide the eye. 100 $\mu$m flux is quite clearly visible associated with the LLCC; the correlation is shown in Figure \ref{irhi}. }
\label{IR}
\end{center}
\end{figure}

To generate a map of the IR flux associated with the LLCC, first we must remove the bulk of the IR flux which is  associated with the background \hia, beyond the local cavity. We make a map of the background \hi column by integrating over the entire Galactic \hi line in the GALFA-\hi data and subtracting the \hi flux associated with the LLCC, as found in \S \ref{ghi_analysis}, model (1). Then we fit a first-order polynomial to the relationship between the background \hi and the IR flux, over the region of the map, using a standard least-squares approach. We have implicitly assumed that the background \hi is largely optically thin, typically a good assumption for high latitude WNM where $T_A \ll T_x$. We then subtract the expected IR background flux from the observed IR flux to determine the residual flux we expect to be associated with the LLCC itself. The result of this process is shown in Figure \ref{IR}.

\begin{figure}
\begin{center}
\includegraphics[scale=0.5, angle=0]{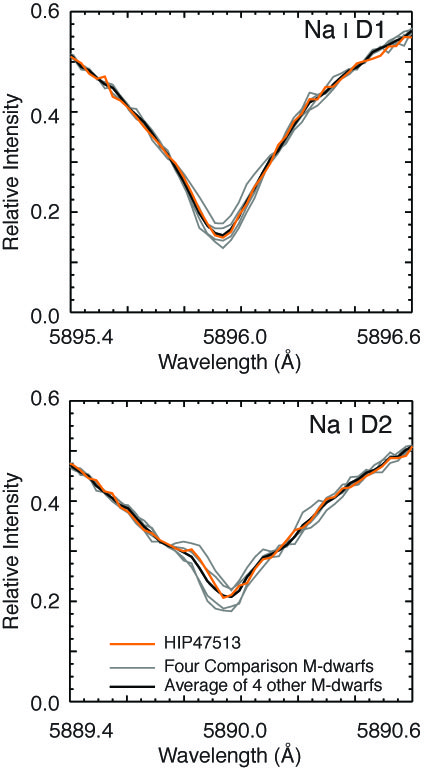}
\caption{The \nai \ $D_{1}$ and \nai \ $D_{2}$ line in \hipstar  \ and the four best-matched CPS stars. The gray lines represent the best-matched stars, the black represents their average, and the orange represents \hipstar. The fact that the \hipstar \ lines nearly exactly match the average of the lines in the four comparison stars indicates there is negligible absorption in \nai \ $D_{1}$ and \nai \ $D_{2}$ from the LLCC toward \hipstar.}
\label{zooms}
\end{center}
\end{figure}

\begin{figure}
\begin{center}
\includegraphics[scale=0.46, angle=0]{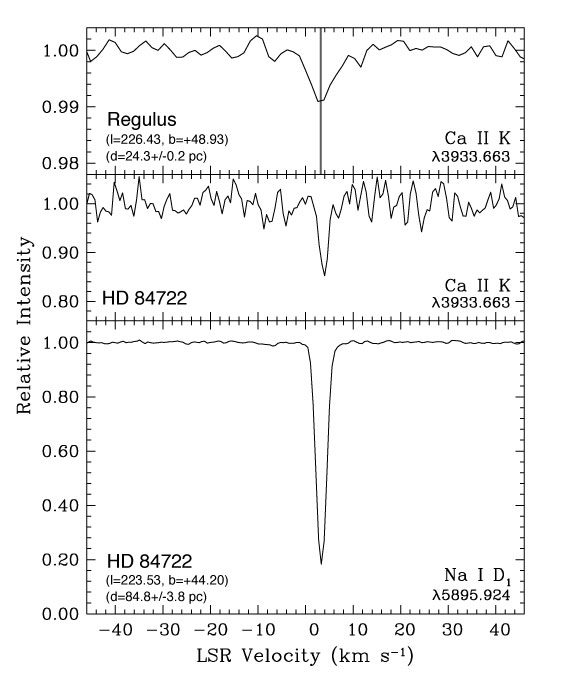}
\caption{The spectrum of Regulus and HD 84722 at the Ca {\sc II} K line. The spectrum of HD 84722 is also shown for the Na {\sc I} $D_1$ line, as a comparison. The velocity of the Regulus cloudlet, 3.19 \kms LSR, is marked with a gray line in the Regulus Ca {\sc II} K spectrum; the near-exact alignment is a clear indication that the Ca {\sc II} K line absorption is created by the Regulus cloudlet. Note that the Ca {\sc II} K line appears broader in the Regulus spectrum because of the somewhat lower spectral resolution of the observation. }
\label{regspect}
\end{center}
\end{figure}

\section{Cloud Properties}\label{cloudprop}

\subsection{Distance}\label{distance}

\subsubsection{Lower Limit}\label{ll}

Ruling out \nai\ absorption along the line of sight to \hipstar \ requires detailed analysis for two reasons. 
First, \hipstar's crowded M-dwarf spectrum hides interstellar absorption 
features far more effectively than smooth A-star spectra like 
HD 84722 in \citet{Meyer:2006dn}. 
Second, and more problematic, is that \hipstar\ has an LSR radial velocity
of +4 km s$^{-1}$ \citep{Nidever02}, indistinguishable from the cloud's 
line-of-sight velocity.
That is to say, the cold cloud's \nai\ absorption features would fall 
right the middle of \hipstar's deep \nai\ D lines.

Because of the difficulty untangling the \nai\ features of the cloud
and the star, we compare the \nai\ D line core of \hipstar\ with the
line cores of four additional stars with similar spectra, selected by the method described in \S \ref{optical}, and shown in Table \ref{keck_table}.
Our goal is to place an upper limit on the \nai between the Sun
and \hipstar.


With four stars in hand, we set about normalizing their spectra to \hipstar. 
In the CPS setup, the \nai\ D lines fall near the edge of one of the 
HIRES echelle orders.
The steep blaze function in that region required a somewhat 
sophisticated normalization procedure. 
Using segments of the spectrum 5 \AA\ from the \nai\ line cores,
we fit a fifth-degree polynomial to both of the \nai\ line wings.

With the \nai\ features normalized across all five spectra, we 
zoomed in on the line cores.
By eye, it appeared that no cold-cloud absorption 
is present in the \hipstar\ spectrum (see Figure \ref{zooms}). 
We next put a quantitative limit on the amount of \nai\
that could be along the line of sight to \hipstar.

First, we averaged the four comparison spectra using a straightforward mean where each spectrum was weighted
equally.
Then, we subtracted the comparison spectrum from the \hipstar\ spectrum
and looked for \nai\ absorption features in the residuals.
By fitting a Gaussian to the residuals at both line cores, we find
a maximum absorption line intensity of $0.00\pm0.15$, equivalent to a \nai\ column density of $0.00 \pm 2.1 \times 10^{10}$ cm$^{-2}$.

In \citet{Meyer:2006dn}, \nai \ column is found in 23 stars behind the LLCC. We find that all 8 stars with measured cold cloud $N_{\rm HI} > 2 \times 10^{19}$ cm$^{-2}$ have $N_{\rm Na I} > 5 \times 10^{11}$ cm$^{-2}$. As the \hi column towards \hipstar \ is $\sim10^{20}$ cm$^{-2}$, the measured 3$\sigma$ limit of $N_{\rm Na I} < 6.3 \times 10^{10}$ cm$^{-2}$ implies conclusively that \hipstar \ lies in front of the LLCC.

\subsubsection{Upper Limit}\label{ul}

As illustrated in Figure \ref{regspect}, our high signal-to-noise
(S/N~$\approx$~900) spectrum of Regulus reveals a weak
interstellar Ca {\sc ii} K line at an LSR velocity  of 3.1 \kmsa.  In order to give some context for the Ca {\sc ii} K line, we have also included spectra of the interstellar \nai\ D$_1$ and
Ca {\sc ii} K absorption toward HD 84722 in Figure \ref{regspect}.  HD 84722 is
positioned at a distance of 84.8~$\pm$~3.8 pc \citep{vanLeeuwen07}
behind the core of the LLCC.  The Na {\sc i} D$_1$ spectrum of HD 84722 is from
\citet{Meyer:2006dn} and the Ca {\sc ii} K spectrum is the product of data
obtained in March 2007 with the KPNO coud\'e feed.  The latter spectrum
was reduced in a similar manner as the Regulus spectrum and it
represents the sum of six T2KB CCD spectra spanning a total exposure
time of 11 hours at a velocity resolution of 1.5~km~s$^{-1}$.  The
measured LSR velocities of the LLCC Ca {\sc ii} K and \nai\ D$_1$ absorption
toward HD 84722 are $+$3.9 and $+$3.7~km~s$^{-1}$, respectively.  The
fact that the \nai\ absorption is much stronger than that of Ca {\sc ii}
toward HD 84722 is consistent with the character of the cold, dense
atomic gas at the core of the LLCC.  

\begin{figure}
\begin{center}
\includegraphics[scale=0.35, angle=0]{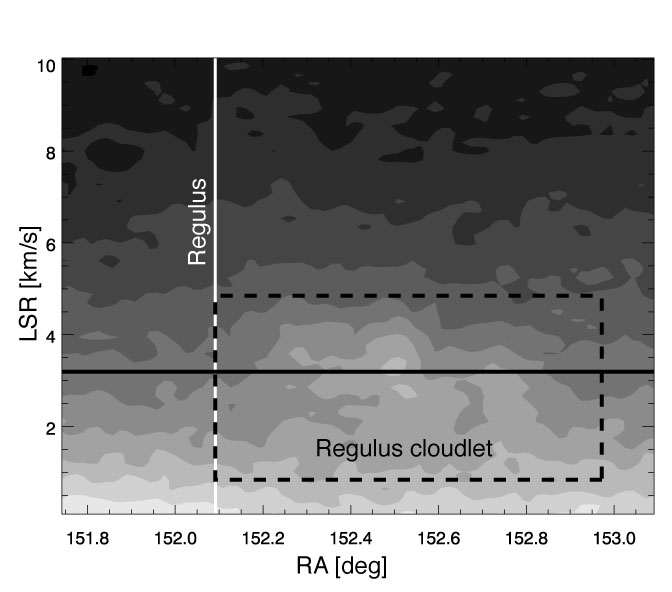}
\caption{An \hi position-velocity slice at $\delta = +11^\circ 58^\prime 30^{\prime\prime}$, including the position of Regulus, marked with a white line. The shading corresponds to brightness temperature of the \hia, with white as higher brightness temperature. The expected velocity (+3.19 \kmsa) of the Regulus cloudlet if it were a part of the LLCC is marked with a black horizontal line, which is also marked in the Regulus spectrum in Figure \ref{regspect}. The position and velocity of the Regulus cloudlet is marked with dashed box. The coincidence of the \hi velocity of the cloudlet, the expected velocity of the cloudlet, and the Ca {\sc II} K line in Regulus indicate that the LLCC is in front of Regulus.}
\label{regcloud}
\end{center}
\end{figure}

Detection of interstellar Ca {\sc II} absorption towards Regulus indicates that some cool interstellar matter must exist between us and Regulus. Unfortunately, Regulus is not along the line of sight directly toward the body of LLCC, but rather about 3$^\circ$ away. Regulus is on the side of the LLCC that seems to have a significant amount of connected cloud debris (see Figure \ref{img_fitnr}), and indeed Regulus is very near a small cloudlet, which we dub the Regulus cloudlet. We expect the Regulus cloudlet is a component of the LLCC for a number of reasons. First, it is only a degree or so away from a long plume of material extending towards higher right ascension from the break between the two main LLCC clouds, and therefore is most likely a fragment of this plume. Secondly, it has a linewidth entirely consistent with the rest of the LLCC. Perhaps most tellingly, when we perform a least-squares fit to the velocities of the main body of the LLCC with a model that accounts for its velocity in 3-space, the velocity expected for the position of the Regulus cloudlet were it to be moving as a solid body with the LLCC is 3.19 \kms LSR. This velocity is in very nice accord with what is observed for the cloudlet (see Figure \ref{regcloud}). This also implies that the Regulus cloudlet is largely dynamically connected to rest of the LLCC, making a scenario in which the cloudlet is at a dramatically different distance rather contrived. 

The next question is whether the absorption in Ca {\sc II} seen in Regulus is associated with Regulus cloudlet. Regulus is on the very margin of the cloudlet, with no obvious cloudlet CNM \hi detectable directly toward the star. The CNM non-detection is consistent with the non-detection of \nai\, typically associated with CNM, as in HD 84722 (Figure \ref{regspect}). Ca {\sc II} is frequently detected towards warmer neutral gas or even largely ionized gas, if the gas is not so hot as to deplete Ca {\sc II} into higher ionization states. Therefore, it is not surprising to detect Ca {\sc II} on the edge of an ablating piece of cloud where material could easily be heating up and mixing with the warmer, more ionized ISM. Perhaps even more conclusively, the Ca {\sc II} absorption in Regulus is at almost exactly the same velocity as the nearest part of the Regulus cloudlet (see Figures \ref{regspect} and \ref{regcloud}). We conclude that we are indeed seeing absorption of Regulus by the cloudlet, and that therefore the LLCC is closer than $24.3 \pm 0.2$ pc. 

\subsection{Temperature}\label{temp}

\citet{CK80} and \citet{HT03} found temperatures for the LLCC ranging from 13 K to 22 K for positions towards strong radio continuum sources. To determine temperatures for the rest of the cloud, we turn to our saturated line fits from \S \ref{ghi_analysis}.

\subsubsection{Deriving the Spin Temperature for Model 5}

We use the standard radiative transfer equation to derive the spin
temperature $T_x$ for the near-side absorbing component. Model 5 consists of
four independent components that contribute to the measured brightness
temperature $T_B(V)$: \begin{enumerate}

\item The continuum background, which consists
of the cosmic microwave background (2.8 K) plus the Galactic continuum
background. We estimate the latter from the 408 MHz \citet{Haslam82} survey by applying a brightness-temperature spectral index of
$-2.6$ to their measured value of $\sim 15$ K; this gives 0.6 K at 1420
MHz. Thus, we take the total continuum brightness temperature to be
$T_{bkgnd} = 3.4$ K.

\item The contribution from background \hia, which we call the WNM
  contribution because the profiles are much wider than those of the
  cold cloud. This is velocity dependent. We represent it by the symbol
  $T_{WNM}(V)$. Numerically, it is represented by the fitted
  polynomial mentioned in \S \ref{ghi_analysis}.

\item The LLCC's more distant optically thin component, which we represent with
  subscript 0. Its brightness temperature in the absence of absorption
  is

\begin{equation} 
 T_{B, 0}(V)= T_0 \exp\left[ - {(V-V_0)^2 \over 2 \, \delta V_0^2} \right]
\end{equation}

\item The closer optically thick component, which we represent with
  subscript 1. Its brightness temperature in the absence of absorption
  is

\begin{mathletters}
\begin{equation} 
T_{B, 1}(V)= T_{x,1} \left(1 - \exp(-\tau_1 (V)) \right)
\end{equation}

\noindent where
\begin{equation} 
\tau_1 (V)= \tau_{1,0} \exp\left[ - {(V-V_1)^2 \over 2 \, \delta V_1^2} \right]
\end{equation}
\end{mathletters}

\end{enumerate}

Putting all these together, the emergent brightness temperature $T_B(V)$
is

\begin{eqnarray}
&& T_B(V) = T_{B, 1}(V) + \\ \nonumber
&& \left[ T_{bkgnd} +  T_{WNM}(V) +  T_{B, 0}(V) \right]\, \exp[-\tau_1 (V)]
\end{eqnarray}

\noindent Our measured profiles are frequency-switched, which means that
we measure the quantity $T_B(V) - T_{bkgnd}$. For each pixel, this is the
fitted function; we derive the spin temperature $T_{x,1}$ from the fitted
parameters.

\begin{figure}
\begin{center}
\includegraphics[scale=0.3]{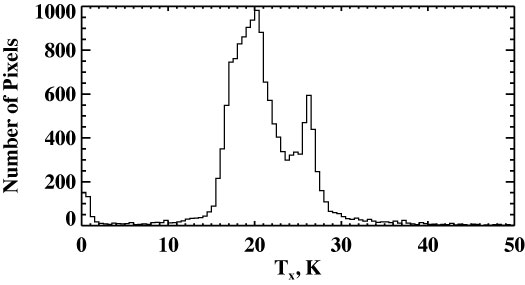}
\end{center}
\caption{A histogram of the derived spin temperature for the front optically
  thick component in Model 5 (see \S \ref{temp}). The cloud is very cold, with modeled temperatures consistent with \hi absorption line observations from \citet{CK80} and \citet{HT03}. 
  \label{plottspin}}
\end{figure}

Figure \ref{plottspin} shows the histogram of derived spin temperatures
$T_{x,1}$ for the near component. It peaks near 20 K and a secondary peak
near 26 K, and is well-confined within the range 15-30 K. Some
anomalously low points with very low spin temperatures are presumably
the result of spurious fits.

\begin{figure}
\begin{center}
\includegraphics[scale=0.3]{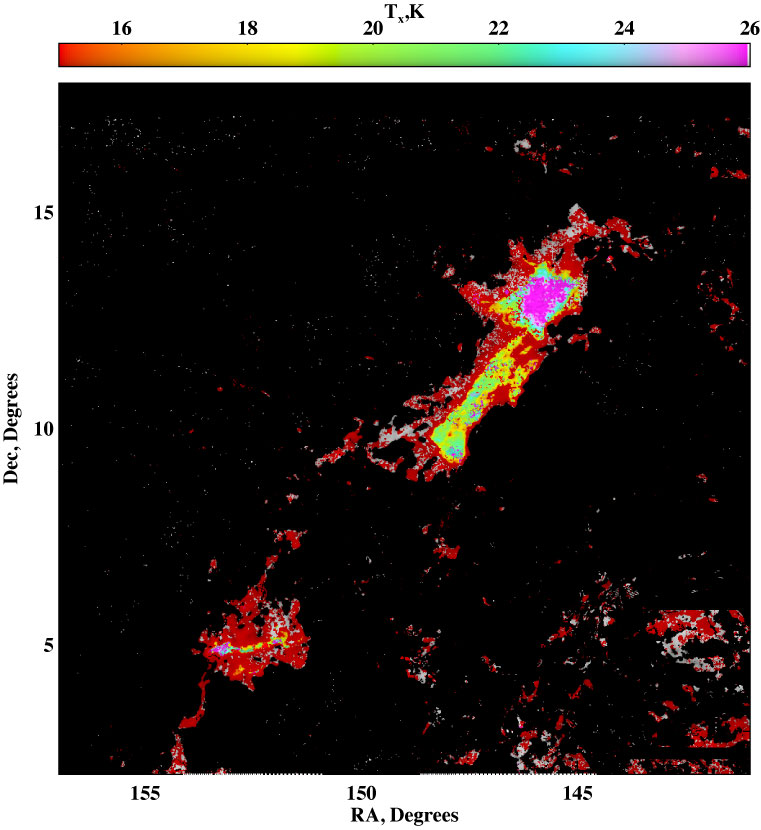}
\end{center}
\caption{ 
 The modeled temperature of the LLCC; color indicates the $T_{x1}$, gray indicates that $T_{x,1}$ is indeterminate.  \label{img_tx}}
\end{figure}

Figure \ref{img_tx} shows the image of $T_{x,1}$. Regions of high column
density tend to be warmer. For weak profiles with low column density,
$T_{x,1}$ is often indeterminate, as shown by the gray.

\subsection{Distance-Dependent Quantities}
Given the lower and upper limits on LLCC distance derived in \S \ref{ll} and \S \ref{ul}, we can determine a range for a variety of other quantities that pertain to the cloud. We determine the mass by simply integrating the \hi column over the area of the cloud, and physical size can be determined from angular size. If we make the very rough assumption that the cloud is as thick as it is wide we can determine a density for the cloud, using a fiducial peak column density of $2.5 \times 10^{20}$ cm$^{-2}$.  Using the temperature range found in \S \ref{temp}, we can determine a range of pressures.  We can determine a rough lifetime for the cloud by dividing the fiducial scale, in this case the cloud width, by the collapse velocity (see \S \ref{kin}). Results are shown in Table 2.

Our assumption that the cloud's thickness along the line of sight is comparable to its width (i.e. the cloud is tubular), is at odds with the result from \citet{HT03} that this cloud is thin along the line of sight (i.e. sheet-like). \citet{HT03} assumed the cloud was outside the local cavity, and thus 4 to 9 times more distant than we now know it to be. This distance assumption gave an overestimate of the width of the cloud and thus an overestimate of its aspect ratio. \citet{HT03} also measured the thickness at low-column ($N_H \simeq 3 \times 10^{19}$ cm$^{-2}$) sight-lines to 3C225a and 3C225b, about 10 times lower column than the thickest part of the cloud in the present analysis. This low column density gave a lower expected thickness for a given measured temperature and assumed pressure, further exacerbating the inconsistency with our work. If the cloud were 10 times flatter along the line of sight than we have assumed, the derived pressures would be 10 times higher, far out of the standard ISM pressure range \citep[e.g.,][]{Wolfire2003}. This result validates our present picture of the cloud as more tubular than sheet-like.

\subsubsection{Kinematics of the Two Velocity Components}\label{kin}

We found that two velocity components usually provide better \hi line fits than a
single velocity component. Model (5) allows us to discriminate between
the velocities of the two components, because the optically thick
foreground component is closer. Accordingly, comparing their velocities
allows us to determine whether the components are approaching or
receding from each other.


There is an overwhelming tendency for the front component to have higher velocities than the
rear. This indicates that the components are approaching one another, typically with
a velocity difference $\sim 0.4$ km s$^{-1}$.

\begin{figure}
\begin{center}
\includegraphics[scale=0.3]{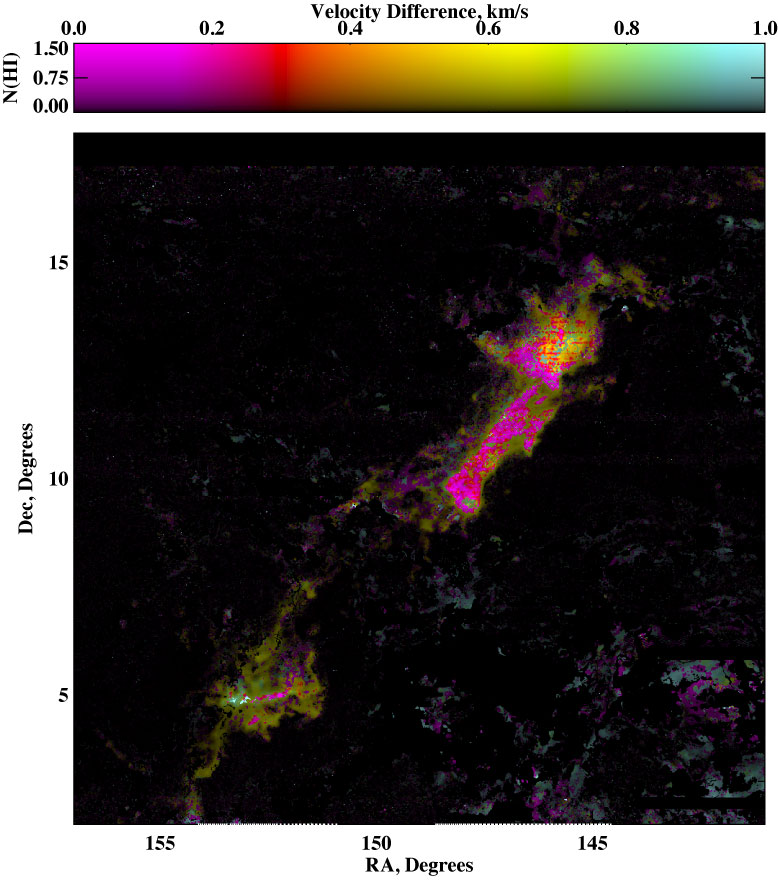}
\end{center}
\caption{A map of the velocity difference between the two colliding components. Brightness is $\propto (T \Delta V)^{0.5}$, summed over both
  velocity components. Color indicates the velocity difference between the two components.  \label{img_veldiff}}
\end{figure}

Figure \ref{img_veldiff} images the velocity difference between the two
components, with color indicating velocity difference and lightness the
total profile area, i.e., the sum of the two component areas. We only
show pixels which had a successful two-component fit, where the two
components were not degenerate. Generally, the edges of the
cloud (where the intensity is low and the line-of-sight thickness is
small) have larger relative velocities.  In these portions of the
cloud, the cloud is getting thinner with time by $\sim 0.4$ pc
Myr$^{-1}$. If we assume the two components are not more separated than the width of the cloud, we find the collision time to be approximately 1 Myr.

\begin{table}[htdp]
\begin{center}
\caption{Quantities for the LLCC and LRCC based on the determined distance range. Density, pressure, and collision time measurements depend on the assumption that the cloud is as thick as it is wide. }
\begin{tabular}{c c c}
quantity & range & unit\\
\hline
distance & 11.3 -- 24.3 & pc\\
LLCC \hi mass &  0.235 -- 1.07 & $M_\odot$\\
LLCC length & 2.8 -- 5.9 & pc \\
LLCC width &  0.25 -- 0.54 & pc \\
LRCC length & 13 -- 28 & pc \\
density &  320 -- 150 & cm$^{-3}$ \\
pressure & 9600 -- 2250 & K cm$^{-3}$ \\
collision time & 0.6 -- 1.3 & Myr \\
\label{ddq}
\end{tabular}

\end{center}
\end{table}

\begin{figure}
\begin{center}
\includegraphics[scale=0.4]{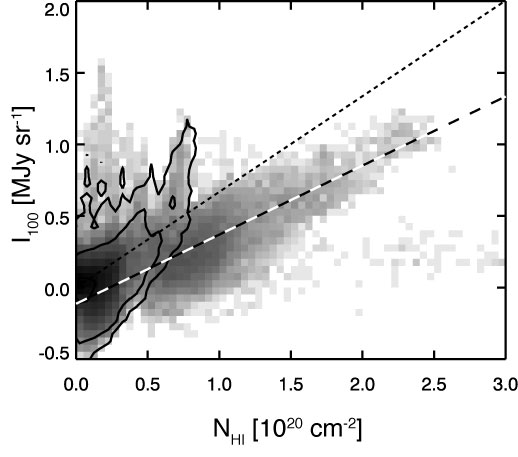}
\end{center}
\caption{ The relationship between \hi column and infrared flux for the LLCC under an \hi optically thin assumption (contours) and the fits in \S \ref{ghi_analysis} (grayscale). The dust-to-gas ratio of 48 $\times$10$^{-22}$ MJy sr$^{-1}$ cm$^{2}$ for the \S \ref{ghi_analysis} fits is shown as a black and white dashed line, while the standard high-latitude dust-to-gas ratio found by \citet{SFD98} of 67 $\times$10$^{-22}$ MJy sr$^{-1}$ cm$^{2}$ is shown in a black dotted line. Note that the optically thin assumption leads to a strange dependence of the dust-to-gas on the \hi column. \label{irhi}}
\end{figure}

\subsection{Dust Content}
To determine the large-grain dust content of the cloud we investigate the relationship between the \hi column density and the 100 $\mu$m emission. Figure \ref{irhi} shows the relationship between these two quantities over the cloud area. It is clear from this plot that there is a very narrow and tight linear relationship between \hi column and IR flux. This serves as an important confirmation that the fitting procedures used in Section \ref{ghi_analysis} to determine the saturated \hi column densities are not wildly incorrect. For comparison we show the relationship between the IR flux and the \hi column under an optically thin assumption, model (1) (Figure \ref{irhi}, contours). These two quantities are clearly not linearly correlated, confirming that an optically thin assumption for the \hi in the LLCC is wrong.

The cloud has a 100 $\mu$m to \hi column value of 48 $\times$10$^{-22}$ MJy sr$^{-1}$ cm$^{2}$. This is somewhat lower than the result from \citet{SFD98} for low column density areas of 67 $\times$10$^{-22}$ MJy sr$^{-1}$ cm$^{2}$. It is also lower than the standard Galactic values found in \citet{Boulanger:1988jr}, about 100 $\times$10$^{-22}$ MJy sr$^{-1}$ cm$^{2}$, although consistent with their measurement towards Auriga. Similarly, it is lower than the average value found for Galactic clouds measured in \citet{Peek:2009up}, but consistent with the values found for the L1, L7, and L8 clouds found in that work. We conclude that this cloud has either a lower-than-expected overall dust grain density, or a population that has somewhat larger grains than typical for the ISM that maintain a somewhat lower temperature. In either case the cloud is only marginally anomalous in IR production via grains. 

\section{Implications for the Local Cavity}\label{ilc}
Authors since Verschuur's discovery paper have remarked upon the observational exceptionality of the LLCC: it is remarkably cold, bright, and long in the Galactic \hi sky. The cloud is also exceptional spatially; it is one of just a few clouds with column densities above $10^{20}$ cm$^{-2}$ within a few dozen pc of the sun (\citei{Vergely10}, RL08). This makes it a excellent and rare test element for studying the contents and dynamics of the gas that surrounds us within the local cavity. 

\subsection{Interaction with the CLIC}
In recent years much pioneering work has been done on the closest known interstellar clouds to the sun \citep[e.g][]{RL08, Frisch08, RF08}. The core results from these investigations is that the sun resides within the CLIC, a collection of clouds within about 15 pc of the sun, and that these CLIC clouds move as solid bodies past the sun in roughly the same direction. These clouds have typical column densities near $10^{18}$ cm$^{-2}$, with temperature near $10^4$ K. Many authors \citep{AH05, V-S06, Heitsch06} have conjectured that CNM clouds are formed by the convergence of flows of warm neutral gas. RL08 pointed out that the LLCC is coincident on the sky with the faster moving Gem cloud and the slower LIC, Auriga, and Leo CLIC clouds, and suggest that the collision of these clouds may be responsible for the formation of the LLCC. 

Our new lower limit on the LLCC of $11.26 \pm 0.21$ pc seems to rule this out. All of the clouds suggested by RL08 to be in collision to form the LLCC are detected to be, at least in part, closer than 11.1 pc (the LIC envelops the sun; Gem is detected within 6.7 pc, Leo within 11.1 pc, and Aur within 3.5 pc). The CLIC clouds certainly span some range of distances, and the extent to which they are entangled is still largely unknown; it is possible that the LLCC exists at some more complex interface between these clouds. That said, the simple picture in which the LLCC is at the interface of two independent colliding clouds seems unrealistic given the new distance lower limit on the LLCC.

\begin{figure}
\begin{center}
\includegraphics[scale=0.5, angle=0]{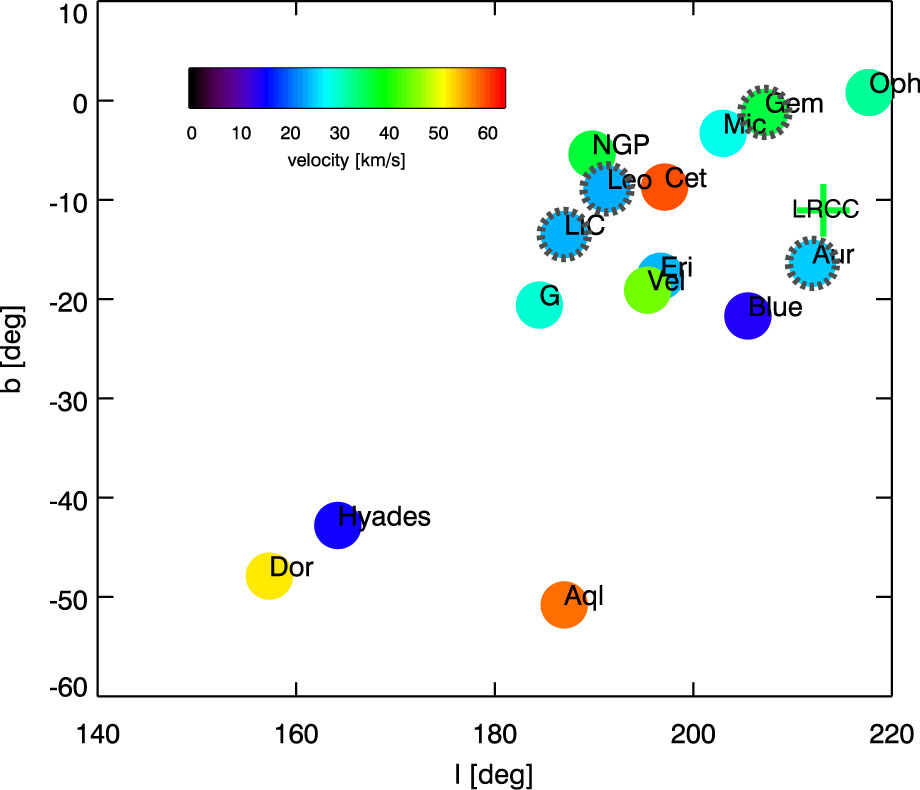}
\caption{A diagram showing the heliocentric velocities of the 15 clouds described in RL08 (circles) and the velocity of the LRCC described in Haud10 (cross). The velocities associated with the 4 clouds closest on the sky to the LRCC are outlined in a dashed gray line. The LRCC has a very similar velocity to the RL08 clouds, suggesting they are associated to each other or some other parent cloud or population.}
\label{velcomp}
\end{center}
\end{figure}

While the distance of the LLCC does seem to separate it somewhat from the CLIC, the velocity of the LLCC may imply that the CLIC and the LLCC are more related. Haud10 fit the LRCC (the ribbon of clouds that contains the LLCC) with a traversing, rotating, expanding, ring. It is unclear if this model has physical significance, as such structures aren't typically found in dynamic models of the ISM, but the results are striking. The overall velocity of the ring is 37.4 \kms toward $l = 213^\circ$, $b = -11^\circ.1$ in the Haud10 fit (when converted to the heliocentric reference frame), which is nicely consistent with the velocities of the CLIC clouds found in RL08 (Figure \ref{velcomp}). It is possible that this consistency is a coincidence, but it seems more likely that the CLIC clouds and the LRCC are related to each other in some way, or to a parent cloud population.

\subsection{Shadowing of Soft X-rays by the LLCC}

As we outlined in the Introduction, there is significant controversy over the origin of soft X-rays seen coming from all directions in the sky. In the local hot bubble theory, soft X-rays at low and intermediate latitudes emanate from a pervasive hot gas filling the local cavity, with the excess of soft X-rays at high latitude coming in part from the halo. In the SWCX theory, the isotropic component of soft X-rays comes from the very nearby \citep[$\sim 100$ AU;][]{Stone05} interaction of the solar wind with the surrounding ISM. We can differentiate between these theories by looking for absorption of these X-rays by the LLCC.

\begin{figure}
\begin{center}
\includegraphics[scale=0.225, angle=0]{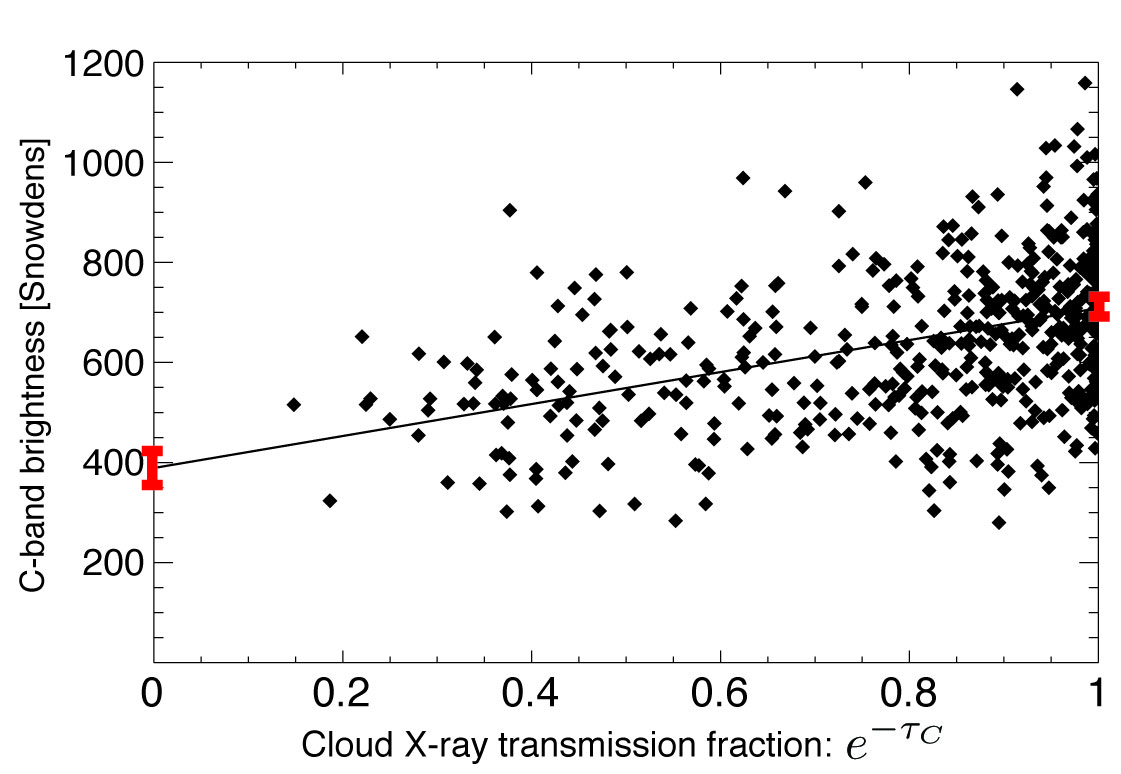}
\caption{The RASS 1/4 keV (C-band) brightness as a function of LLCC transparency in the C-band, given the measured \hi column. Diamonds represent pixels in the RASS map, and the line represents the fit found in Equation \ref{atten_eqn}. The red bars represent 1 $\sigma$ errors in $I_{C, foreground}$ and $I_{C, total}$. The fact that $398 \pm 38$ Snowdens emanate from in front of the cloud rules out an model with no SWCX and a smoothly emitting LHB (see \S \ref{moda}). \label{tauvrosat}}
\end{center}
\end{figure}

In the direction of the LLCC the wall of the local cavity is between 100 pc and 150 pc away \citep{Meyer:2006dn}, and we have demonstrated that the LLCC is between 11.26 pc and 24.3 pc away (\S \ref{distance}). We have determined the \hi column of the LLCC rather carefully (\S \ref{ghi_analysis}), and the local bubble wall has a column density of $\sim 3 \times 10^{20}$ cm$^{-2}$ in the direction to the LLCC. For a reasonable spectral energy distribution over the 1/4 keV band of ROSAT, an optical depth of 1 corresponds to an \hi column of $1.05 \times 10^{20}$ cm$^{-2}$ \citep{Snowden97}.
To determine the amount of flux coming from in front of the LLCC we fit the equation
\beq\label{atten_eqn}
I_C\left(\alpha, \delta\right) = I_{C, foreground} + I_{C, background} e^{-\tau_C\left(HI\left(\alpha, \delta\right)\right)}
\eeq
where $I_C\left(\alpha, \delta \right)$ is the flux detected toward the cloud in the SXRB 1/4 keV (C-band) map, and $\tau_C\left(HI\left(\alpha, \delta \right)\right)$ is the optical depth to C-band X-rays caused by the LLCC as a function of position on the cloud.  $I_{C, foreground}$ and $I_{C, background}$ are the unknown X-ray flux from in front of and behind the cloud respectively. To do this we use a standard least-squares fit. The standard errors derived from a least-squares fit are correct under the assumption that the deviation from the model is Gaussian and uncorrelated. Neither of these assumptions is true in our case, since there are obvious structures in the SXRB 1/4 keV maps in this direction that are far from random, and will generate strong, unmodeled errors. To account for this we use the displacement mapping technique described in \citet{Peek:2009up}, which determines the errors by examining the fluctuation in the result as the cloud is placed in different positions on the nearby sky. We then use this fluctuation to better estimate the errors. Using this method we find $I_{C, total} = I_{C, background} + I_{C, foreground} = 708 \pm 11$ Snowdens and $ I_{C, foreground} = 398 \pm 38$ Snowdens (see Figure \ref{tauvrosat}).

To constrain models for the origins of the soft X-rays, we combine the above result with the results toward the intermediate latitude ($b = -34.04$) molecular cloud MBM 12 in \citet{1997ApJ...484..245K} and \citet{KS2000}. In this analysis the authors found that MBM 12 was at the back of the local bubble, as it shows very little, if any, shadowing in the C-band, is at a distance of $\sim$ 90 pc \citep{1999A&A...346..785S}, and with a C-band flux of 347$^{+5}_{-10}$. Any model we construct for the origin of the C-band X-rays must account for these two measurements.

\subsubsection{Model A: Local Hot Bubble and Halo Emission}\label{moda}

In this model the LHB is filled with a homogeneous gas such that the intensity, $I_C$, is simply proportional to the path length of LHB gas. We add to this some unabsorbed halo emission from the halo, equivalent to the trans-absorptive emission discussed in \citet{KS2000}. We can show easily that model A is insufficient to explain the data. We find that $389 \pm 38$ Snowdens of soft X-ray flux are not attenuated at all by the LLCC, and thus are generated closer to us than the LLCC, along a path length no longer than 25 pc. If these X-rays were generated by a hot, local cavity--filling gas, we would expect at most 708 Snowdens *(25 pc / 100 pc) = 177 Snowdens to originate from in front of the cloud, as the ratio of path lengths is no more than 25 pc / 100 pc. Adding any halo emission only exacerbates the incongruence. This expected value of 177 Snowdens is inconsistent with our result by more than 5 $\sigma$, thus strictly ruling out model A. 

\subsubsection{Model B: Solar Wind Charge Exchange and Halo Emission}

In model B we assume the LHB does not contribute at all to the soft X-ray background. Instead we appeal to solar wind charge exchange, which we model as a single isotropic, all-sky intensity in the C-band. We note that models of SWCX by \citep{Koutroumpa08} find an expected maximum dipolar variation of $\sim$25\% in the all-sky C-band maps from the direction of motion of the heliopause through the LIC, although they do not model the outer heliospheric region, which may largely mitigate this variation. An all-sky value of 347 Snowdens for the SWCX is 1.1 $\sigma$ off from our measurement of the C-band intensity in front of the LLCC, and also satisfies the intensity measurement toward MBM 12. The background intensity off the LLCC of $708 - 389 = 319$ Snowdens is explained by the halo emission. We note that if the background flux could as easily be described by the ``Hot Top'' framework of \citet{WS10} as by halo emission. In either case, model B is consistent with these results.

\subsubsection{Model C: Local Hot Bubble, Solar Wind Charge Exchange, and Halo Emission}

In model C we allow all three components to contribute to the soft X-ray background. Since model B fits the data successfully, we concern ourselves with determining the maximal LHB contribution that is consistent with the data. To have a non-zero contribution from the LHB, the true value towards the LLCC must be less than the true value toward MBM 12 (as MBM 12 is more distant), which is 1.1 $\sigma$ from our measurement. If we take 3 $\sigma$ as a credible limit, we find that the LHB has a maximal emission of 1.1 Snowdens / pc, about 4 times lower than found in \citet{KS2000}, with 247 Snowdens coming from SWCX. At 2 $\sigma$, the emission is only 0.52 Snowdens / pc, with 300 Snowdens coming from SWCX. 

\section{Conclusion}\label{conc}
We have examined the the LLCC using \hi emission data, diffuse far-infrared dust emission data, optical stellar absorption line data, and soft X-ray background data. We have concluded that the LLCC is within 24.3 pc (Regulus), but beyond 11.3 pc (HIP 47513). Through careful fitting of the \hi line, we have found that the cloud is typically composed of two components, moving toward each other at 0.4 \kmsa. We also have found typical temperatures for the LLCC between 15 K and 30 K, with column densities ranging up to $2.5 \times 10^{20}$ cm$^{-2}$. We have found that the LLCC is too far away to be created by the collisions of the CLIC clouds discussed in RL08, but that the velocity determined for the parent LRCC cloud is very consistent with the CLIC clouds, pointing to a common origin. Perhaps most importantly, we have shown that the extent to which the LLCC absorbs C-band radiation is inconsistent with a homogeneous Local Hot Bubble interpretation for the isotropic soft X-ray background. We have shown that this absorption instead favors the solar wind charge exchange interpretation, though we do not strictly rule out a model in which both contribute the the soft X-ray background.

While we have answered many questions in this work, we have also posed many question. Is the multi-peaked profile a standard aspect of ultra-cold CNM? Is there significant variability in the SXRB emission within 25 pc? What is the chemical composition of the LRCC? We hope to address these questions with new observations of the LRCC. We intend to complete the map of the LRCC above $\delta = -1^\circ$ with the GALFA-\hi survey, using archival and scheduled observations. These observations will allow us to further constrain the SXRB emanating from within the distance to the cloud, as some regions appear have high column density similar to the LLCC. The larger cloud area to examine may also allow us to narrow the distance range with new and archival high resolution stellar absorption observations. Observations are planned of various metal lines in the optical and UV towards stars behind the LRCC, which will allow us to better understand the physical state of the cloud, including its metallicity, ionization fraction, and pressure.

Acknowledgements: JEGP would like to thank Barry Welsh, Seth Redfield, and Mordecai Mac Low for many helpful conversations. KMGP would like to thank Geoffrey Marcy for helpful conversations and Deborah Fischer, Steve Vogt, Paul Butler for their generosity with the CPS data. The authors would like to thank the referee, Jeffrey Linsky, for detailed comments which greatly improved the manuscript. This research was funded in part by National Science Foundation grants AST07-09347, AST09-08841, and AST- 0917810.

\appendix

We handled the ``guessing problem'' mentioned in \S \ref{fitgau} using the following sequence for
each pixel: \\

\begin{enumerate}

\item We first fit model (1), the unsaturated single Gaussian. We
  considered the velocity range --4.0 to +10.0 \kms and defined the
  ``line'' velocity range as --1.2 to +7.5 \kmsa. To define the guesses,
  we first subtracted a third-order polynomial fit to the off-line
  profile values. We took the height guess equal to the maximum profile
  value within the line velocity range. We took the guesses for center,
  and width (our widths are always FWHMs) equal to the first and second
  moments of the HI profile in the line range. Finally, using the
  results of this fit, we fit simultaneously for the polynomial and the
  Gaussian parameters over the full velocity range; this fit was almost
  always successful, and when it wasn't we used the results from the
  first fit.

\item We use the solutions from step 1 to generate guesses for
  model (2). Guesses for the two centers are equal to the
  model (1) fitted center $\pm X$, where $X= 0.3$ times the model (1)
  fitted width; for the widths, we always used 0.8 km s$^{-1}$; and both
  heights were equal to model (1) fitted height. 

\item We use these solutions from step 2 to generate guesses for
  model (3). The guesses for centers and widths are equal to
  the solutions, the guesses for optical depths are both equal to 1.5,
  and the guesses for spin temperature are adjusted to make the
  unabsorbed peak brightness temperature of the components' emission
  lines equal to their model (2) peak brightness temperatures. The
  guesses for model (5) are obtained in a similar fashion.

\item The guesses for models (4) are obtained in a similar
  fashion as for model (3), except that for $T_x$ and $\tau$ we
  performed fits over a closely spaced grid of $T_x$, so no guessing was
  involved for these two parameters.

\end{enumerate}

\bibliographystyle{apj}

\end{document}